# Parallelization of an implicit algorithm for multi-dimensional particle-in-cell simulations

G. M. Petrov and J. Davis

Naval Research Laboratory, Plasma Physics Division,

4555 Overlook Ave. SW, Washington, DC 20375, USA

**Abstract**

The implicit 2D3V particle-in-cell (PIC) code developed to study the interaction of ultrashort pulse lasers with matter [G. M. Petrov and J. Davis, Computer Phys. Comm. **179**, 868 (2008); Phys. Plasmas 18, 073102 (2011)] has been parallelized using MPI (Message Passing Interface). Details on the algorithm implementation are given with emphasis on code optimization by overlapping computations with communications. Performance evaluation has been made on a small Linux cluster with 32 processors for two typical regimes of PIC operation: "particle dominated", for which the bulk of the computation time is spent on pushing particles, and "field dominated", for which computing the fields is prevalent. We found that the MPI implementation of the code offers a significant numerical speedup. In the "particle dominated" regime it is close to the maximum theoretical one, while in the other regime it is about 75–80 % of the maximum speed-up. The code parallelization will allow future implementation of atomic physics and extension to three dimensions.

*PACS:* 52.38.-r, 52.38.Ph, 52.50.Jm

*Keywords:* Particle-in-cell, Maxwell equations, MPI, Laser–target interaction

## I. Introduction

The particle-in-cell (PIC) codes are ubiquitous and have many applications covering diverse scientific areas such as astrophysics, plasma physics, microelectronics and chemistry [1,2,3]. PIC codes are also at the forefront of simulation tools for modeling laser-matter interactions since they can adequately model both the laser radiation and the response of the material allowing a self-consistent description of particles and fields. One example is the interaction of short-pulse lasers with thin foils, which are routinely used in laboratories around the world for particle acceleration, x-ray generation, and other scientific endeavors such as laser nuclear physics.

The modeling of short-pulse lasers interacting with thin foils faces a host of challenges. Typically, the electromagnetic fields of the laser exceed the atomic field strength and the material becomes instantaneously ionized by Optical Field Ionization. The resulting dense plasma may be opaque for the laser radiation and the material starts to act as a "mirror". The electromagnetic fields can not penetrate the plasma they created and decay inside the material on a scale length of only 10–50 nm, which can be up to three orders of magnitude smaller than the thickness of the foil. In addition, the spatial distribution of the material is highly non-uniform: the target is very dense and concentrated in a narrow region of the computational domain (see Figure 1), contrary to other problems in which the plasma is distributed uniformly [3]. As a result, the numerical solution of the problem requires high spatial resolution and the number of particles is often prohibitively large. The problem is pushed to the extreme in virtually every numerical aspect, which makes the simulations computationally intensive and numerical algorithms challenging. PIC codes can run for days, sometimes weeks, which is a great impetus for development of efficient and robust PIC codes.

In terms of numerical approach, PIC codes fall into two general categories: explicit and implicit. In explicit PIC information (particle positions and fields) is used only from previous time levels for physical quantities that are already calculated, which makes it straightforward and computationally efficient (per time step). It is, however, subject to severe numerical stability constraints [4]. In contrast, implicit PIC codes include information from the next time level, which is more involved and requires some extra logistic and programming efforts, but the payoff can be substantial: the numerical stability improves dramatically, and the number of particles, temporal and spatial resolution requirements are greatly relaxed. Implicit PIC codes with application to laser-target interactions first appeared in the early 80's [5,6]. Since then, numerous codes have been developed over the years: ANTHEM [7] AVANTI [8,9], MACROS [10], DADIPIC [11], OSIRIS [12], LSP [13,14,] CELESTE3D [15], and iPIC3D [4]. Many of those codes are widely used and are constantly evolving both in terms of applications and numerical implementation. Markidis *et al* proposed muti-scale simulations with large dynamic range for studying phenomena spanning over large time scale [4]. Advanced implicit PIC codes have emerged incorporating novel adaptive techniques [16] and adding critical features such as energy conserving schemes [17].

The implicit PIC code, developed at the Naval Research Laboratory (NRL) [18,19] for studying laser-matter interactions, is powerful enough to handle real-world problems with sufficient accuracy within reasonable period of time (hours to a few days). But in spite of the improvements, the code is still unable to tackle some large-scale problems. As an example, if the mass of the target is too large [20], a huge number of particles is required, which is beyond the capabilities of a single processor machine. This problem is unlikely to go away in the near future due to stagnation in CPU speed. The next logical step is code parallelization, i.e. using not one, but many computing cores.

Code parallelization requires careful planning. To achieve maximum efficiency the computational work must be properly distributed among the processors overlapping computations with communications. Parallelizing implicit PIC is less straightforward than explicit PIC, because the former is more complex. In explicit PIC all quantities are calculated sequentially in time, while in implicit PIC source terms are predicted or evaluated at a future time level. Unlike its explicit counterpart, the current density is not computed and can not be used directly in the Maxwell equations. One of the most challenging aspects of implicit PIC codes stems from the fact that they follow different numerical schemes, which entails different parallelization strategies. For example, iPIC3D solves a second order partial differential equation (PDE) for the electric field components (the wave equation), while we solve two first order PDE's for the electric and magnetic field. In addition, in our code the electric field components are coupled and the corresponding equations must be solved simultaneously (see Section 2). The sequence for pushing particles and computing fields is subject to variations too. In LSP [13,14] particles are pushed twice per time step and in iPIC3D [4] an iterative procedure for the particle pusher is employed, while in our code the particles are pushed only once. The computational domain decomposition is also different with OSIRIS and iPIC3D opting for more advanced domain decomposition schemes. The choice of communications between processors is critical for the performance of the (parallelized) code. While some implicit PIC codes have been parallelized using blocking communications [4], we chose non-blocking communications. Since the parallelization is highly dependent on the specific numerical implementation, the parallelization routine is unique for each approach.

It is the purpose of this paper to present a step-by-step parallelization routine of the implicit

PIC code developed in Refs. [18,19]. We employed the most popular approach for code parallelization, the Message Passing Interface (MPI). The parallelized code is written in C++ using Open MPI [21]. The computational speedup is evaluated on a small scale Linux cluster system.

**2. The implicit PIC method.**

In this section we briefly reiterate the implicit PIC method developed in Refs. [18,19]. The Maxwell equations

$$\frac{\partial \vec{E}}{\partial t} = \frac{1}{\varepsilon_0}\left(\vec{\nabla} \times \vec{H} - \vec{j}\right) \tag{1a}$$

$$\frac{\partial \vec{H}}{\partial t} = -\frac{1}{\mu_0}\vec{\nabla} \times \vec{E} \tag{1b}$$

with field components $\vec{E} = (E_x, E_y, E_z)$ $\vec{H} = (H_x, H_y, H_z)$ and $\vec{B} = \mu_0 \vec{H}$ are solved in a Cartesian coordinate system with $\vec{j} = (j_x, j_y, j_z)$ being the conduction current density. The plasma is modeled by a set of relativistic equations of motion

$$\frac{d\vec{p}_\alpha}{dt} = q_\alpha\left(\vec{E}_\alpha + \frac{\vec{p}_\alpha}{m_\alpha \gamma_\alpha} \times \vec{B}_\alpha\right) \tag{2a}$$

$$\vec{v}_\alpha = \frac{\vec{p}_\alpha/m_\alpha}{\sqrt{1+(\vec{p}_\alpha/m_\alpha c)^2}} \tag{2b}$$

$$\frac{d\vec{r}_\alpha}{dt} = \vec{v}_\alpha \tag{2c}$$

for each computational particle α immersed in an electromagnetic field. In Eq. (2) $\vec{r}_\alpha = (x_\alpha, y_\alpha, z_\alpha)$, $\vec{p}_\alpha = (p_{\alpha,x}, p_{\alpha,y}, p_{\alpha,z})$ and $\vec{v}_\alpha = (v_{\alpha,x}, v_{\alpha,y}, v_{\alpha,z})$, are the radius vector, particle relativistic momentum and velocity of computational particle α, $m_\alpha$, $q_\alpha$ and $n_\alpha$ are the mass, charge and the density carried by the particle, respectively, $\gamma_\alpha = \sqrt{1+(\vec{p}_\alpha/m_\alpha c)^2}$ is the relativistic factor, $c$ is the speed of light and $\vec{E}_\alpha$ and $\vec{B}_\alpha$ are the electric and magnetic field at the particle position $\vec{r}_\alpha$. The Maxwell equations (1) are coupled to the particle equations of motion (2) through the conduction current density $\vec{j}(\vec{r}) = \sum_\alpha \vec{j}_\alpha(\vec{r}_\alpha)W(\vec{r}-\vec{r}_\alpha)$, where $\vec{j}_\alpha = n_\alpha q_\alpha \vec{v}_\alpha$ is the current density carried by computational particle α and $W$ is the particle shape function, which is used to distribute quantities from the particle position onto grid nodes and vice versa. The coupling of $\vec{j}$ to the Maxwell equations (1) is critical for the performance of the PIC code. In explicit codes the current density is computed directly and the result inserted into Eq. (1a). In our implicit algorithm Equation (2a) is inverted and solved for the particle momentum. During the process the electric field is factored out and $\vec{j}(\vec{r})$ is put into the form $\vec{j}(\vec{r}) = \sum_{\alpha} (\hat{S}_{\alpha} \vec{E}_{\alpha} + \delta \vec{j}_{\alpha}) W(\vec{r} - \vec{r}_{\alpha})$, which in vector notations reads:

$$\vec{j} = \hat{S}\vec{E} + \delta\vec{j}. \qquad (3)$$

In Equation (3) $\hat{S}$ is a global tensor and $\delta\vec{j}$ is some residual current, both accumulated on grid notes. Their full derivation is given in Ref. [19]. Inserting (3) into the Maxwell equations yields a system of three coupled equations for the electric field components $\vec{E} = (E_x, E_y, E_z)$:

$$(\hat{I} + \hat{S}^{n+1/2})\vec{E}^{n+1} = (\hat{I} - \hat{S}^{n+1/2})\vec{E}^n + (\vec{\nabla} \times \vec{H}^{n+1/2} - \delta\vec{j})\Delta t / \varepsilon_0. \qquad (4)$$

which are solved simultaneously. The magnetic field components are advanced according to

$$\vec{H}^{n+3/2} = \vec{H}^{n+1/2} - \vec{\nabla} \times \vec{E}^{n+1} \Delta t / \mu_0, \qquad (5)$$

once the electric field components $\vec{E}^{n+1}$ are calculated. We would like to point out that the arrangement of electric and magnetic fields on the computational cell does not follow the conventional Yee scheme. The electric field is located on grid nodes, while the magnetic field is located in cell center instead. The next section will elaborate on the parallelization of the implicit algorithm.

**3. MPI implementation of the implicit algorithm.**

The code parallelization is based on standard MPI libraries. For maximum efficiency only non-blocking parallel communications among processors are used. In MPI the computational work is distributed among many processors. Code development is somewhat more complicated since computations flow in parallel on all processors simultaneously and need to be well coordinated in order to avoid bottlenecks and delays. Efficient MPI parallelization is achieved by following a few simple rules such as:

- *Load balancing.* The computational load must be approximately equally distributed among the processors. Specifically for PIC, the electromagnetic fields are recomputed and the particles are pushed for every computational cycle, therefore, the processors must be given equal slice of the computational domain and the particles must be evenly distributed.

- *Overlapping communications with computations.* Since communications are costly, good programming practice requires that some computations are done while data are "in transit" from one processor to another.

Though more sophisticated methods are needed to take full advantage of MPI, we find that these two are sufficient for small scale implementation (tens to hundreds of processors). In the following sub-sections the individual steps of the parallelized PIC algorithm are described. They are adapted specifically for our implicit algorithm, but other PIC codes (including explicit) can follow a similar routine with minor modifications.

3.1 Computational domain decomposition

We employ a 2D Cartesian geometry as shown in Fig. 1. The computational domain is defined as $R = \{0 \leq x \leq L_x, 0 \leq y \leq L_y\}$ with $L_x$ and $L_y$ being its length and width, respectively. The laser electromagnetic radiation advances from left to right along the "x" axis. At the beginning of the computations the front of the electromagnetic pulse is at the "left" boundary $x = 0$. The foil is perpendicular to the laser beam and has a width $W$ and length $L$. The domain decomposition is shown in Fig. 1. The local computational domains are long narrow strips parallel to the "x" axis. Though there are more efficient ways, such as 2D domain decomposition [21], this partition is simple and load balancing for both particles and fields is accomplished. The boundaries between processors are "sharp", i.e. two adjacent processors have common grid nodes but no common cells (Fig. 1). Particles are divided among processors according to their location and transferred to another processor if exiting the local processor computational domain.

3.2 Initialization

The PIC cycle is preceded by an initialization routine, which includes:

(1) Set up a grid on local processors according to the space decomposition. We set a "global" grid and assign each processor a slice of it, forming a local grid with $(m+1) \times (n+1)$ grid points ($n$ is along axis "y" and is not necessarily equal for all processors).

(2) Distribute particles among processors according to their location and initialize them by assigning (local) coordinates, velocities, charge, etc.

(3) Initialize the electromagnetic fields. The electric field components are located on grid nodes, $\vec{E}_{i,j}$, $0 \leq i \leq n, 0 \leq j \leq m$, while the magnetic fields components $\vec{H}_{i,j}$, $0 \leq i \leq n-1, 0 \leq j \leq m-1$ are located in cell center [19]. At the beginning of the computations the electromagnetic field components inside the computational domain of every processor are set to zero.

3.3 The PIC cycle.

The PIC cycle consists of electromagnetic field solver, computation of forces, particle pusher and current density computation followed by interpolation on grid nodes. The electromagnetic field solver is broken into nine steps (Fig. 2). The motivation for having so many steps is dictated by the desire to achieve maximum speed and efficiency of the code. The main hurdle is the necessity to exchange information between processors, which arises because generally speaking the spatial derivatives of $rot(\vec{E})$ and $rot(\vec{H})$ may require information stored on adjacent processors. Since the information exchange is much slower than floating point computations, computational speedup is achieved by carefully overlapping communications with computations. The nine steps, described below, are designed to optimize this process by initiating communications first, performing computations on parts of the computational domain that are independent of the data being sent and finally, receiving the data. For this purpose, we use non-blocking communications between processors. The following steps describe the order in which $\vec{E}_{i,j}$ and $\vec{H}_{i,j}$ are computed. The equation number is that of Ref. [19].

*Step 1:* Compute tensor $\hat{S}$ and vector $\delta\vec{j}$ on grid nodes on each processor (Eq. 12 [19]). On the boundary between two processors the values on grid nodes are incomplete since there is a contribution from particles belonging to the adjacent processor. Non-blocking communication is initiated, sending the first row of local values $\hat{S}_{0,j}$ and $\delta\vec{j}_{0,j}$, $0 \leq j \leq m$ from processor $k+1$ to processor $k$. Total of twelve rows are sent, nine for $\hat{S}$ and three for $\delta\vec{j}$. Later, the data will be received by processor $k$ and added to the last row of local values $\hat{S}_{n,j}$ and $\delta\vec{j}_{n,j}$, $0 \leq j \leq m$.

*Step 2:* In order to compute the electric field on the last row *n* of processor *k*, $\vec{E}_{n,j}$, $0 \leq j \leq m$, one needs to know the local values of the magnetic field $\vec{H}_{n,j}$. However, the data for the magnetic field on local processor *k* span only up to row $n-1$. The needed row *n* belongs to the next processor *k+1* as row 0. Therefore, we initiate a non-blocking communication by sending the first row of the magnetic field $\vec{H}_{0,j}$, $0 \leq j \leq m-1$ from processor $k+1$ to processor $k$. Later, the data will be received by processor *k* and stored as local values $\vec{H}_{n,j}$, $0 \leq j \leq m-1$.

*Step 3:* Apply boundary conditions at $x = 0$ and $x = L_x$ (Eq. 14 [19]) to update $\vec{E}_{i,0}$ and $\vec{E}_{i,m}$, $0 \leq i \leq n$ on local processor *k*.

*Step 4:* Add a source field at $x = 0$ (Eq. 15) to update $\vec{E}_{i,0}$, $0 \leq i \leq n$ on local processor *k*.

*Step 5:* Compute the interior points for the electric field on local processor *k*, $\vec{E}_{i,j}$, $1 \leq i \leq n, 1 \leq j \leq m$.

*Step 6:* The electric field values $\vec{E}_{n,j}$, $0 \leq j \leq m$ belonging to the last row *n* are computed next. On the last processor apply boundary condition (Eq. 14). On all other processors wait (if necessary) for the communications initiated in steps (1) and (2) to complete, retrieve $\hat{S}_{n,j}$ and $\delta \vec{j}_{n,j}$, $0 \leq j \leq m$, as well as $\vec{H}_{n,j}$, $0 \leq j \leq m-1$ and compute $\vec{E}_{n,j}$.

*Step 7:* The electric field values belonging to the first row $\vec{E}_{0,j}$, $0 \leq j \leq m$ on processor $k$ are still not known, but since for $k \geq 1$ they are identical to the electric field values of the last row $n$ of processor $k-1$ (shared grid nodes), initiate a non-blocking communication by sending the last row of values $\vec{E}_{n,j}$, $0 \leq j \leq m$ computed in step (6) from processor $k-1$ to processor $k$. Later, the data will be received by processor $k$ and stored as local values $\vec{E}_{0,j}$, $0 \leq j \leq m$. For local processor $k = 0$ apply boundary condition (Eq. 14).

*Step 8:* While data sent in step (7) are in transit, compute the magnetic field $\vec{H}_{i,j}$, $1 \leq i \leq n-1, 0 \leq j \leq m-1$, using Eq. (13 [19]) for all rows except the first one, $i = 0$. This row can not be computed until the data sent in step (7) are received and become available. Note that no boundary conditions are required for the magnetic field components since all values are computed using the values of the electric field.

*Step 9:* Wait (if necessary) for communications initiated in step (7) to complete and compute the magnetic field $\vec{H}_{0,j}$, $0 \leq j \leq m-1$ on all processors.

The total communication cost is $12m$ for the coefficients forming current density (3), $3(m-1)$ for the magnetic field and $3m$ for the electric field, or the total of $18m$ double precision numbers (eighteen rows), which is about twice the amount of data transfer for explicit PIC codes. However, the overlap of communications with computations compensates to a large extent for the extra communication cost. For example, steps (3)–(5) can be performed while the data sent on steps (1) and (2) are in transit. There is plenty of data to be computed, on the order of $(m+1) \times n$ grid points (all internal grid nodes plus boundary conditions), without any need of the data currently in transit. The same holds for the magnetic field computation in steps (8) and (9). Overall, the procedure is very efficient and is expected to scale well with the number of processors.

The particle pusher is the same as previously described in Ref. [19], but one has to account for particles that cross the boundary with adjacent processors. Those that do are put in a separate array, removed from the current processor and the array of particles is send to the corresponding processor. Actually, except for the first and last processors, two such arrays are formed, one sent from processor $k$ to processor $k+1$ and the other from processor $k$ to processor $k-1$. While the data are in transit, more computations can be done such as collisions or ionizations.

A major issue faced in all PIC codes is the interpolation of quantities from grid nodes to the particle position. The particle shape function, $W$, is a polynomial, often up to fourth order in order to reduce spurious grid heating. The higher the order, the more cells are involved, which presents a problem during parallelization since some of the cells may reside on a neighboring processor. Using "ghost cells" residing on adjacent processors is the most common technique. In the implementation outlined above the use of "ghost cells" is completely avoided since in our implicit PIC linear interpolation yields sufficient accuracy and all quantities are collected within a cell using the procedure outlined in Ref. [3] (Eq. 49-56).

## 4. Results and discussions.

Our first task was to confirm that any multiprocessor run reproduces the single processor one, i.e. several sets of results from the "original code" were reproduced. Once benchmarking was successfully completed, we evaluated the system speedup using two "standard" simulation runs. The PIC code was run on a single processor and repeated on a small Linux cluster with 32 processors. The simulation box is a square with dimensions $\{L_x \times L_y\} = 30 \times 30$ μm$^2$. The number of cells is $1200 \times 1200$ and cell size is $25 \times 25$ nm$^2$. Linearly polarized laser pulses with peak intensity $I_0 = 2.5 \times 10^{24}$ W/m$^2$, pulse duration $\tau_{FWHM} = 60$ fs, spot size $D_{FWHM} = 3.3$ μm, and wavelength $\lambda_0 = 1$ μm are used. The two targets are carbon foils with density $\rho = 1000$ kg/m$^3$, thickness $L_C = 1$ μm (target #1) and $L_C = 0.2$ μm (target #2) and width $W = 28$ μm, with a thin H$_2$O contamination layer on the back with density $\rho = 1000$ kg/m$^3$ and thickness $L_{H_2O} = 5$ nm. It is located 3 μm from the "left" boundary. The particles and electromagnetic field components are advanced with a time step $\Delta t = 0.01 \lambda_0 / c$. The simulations begin at time $t = -10$ fs when the laser pulse enters the computational domain at $x = 0$ and continue for another 180 fs. At the beginning on the simulations the number of computational particles is $2 \times 10^6$ for target #1 and $0.4 \times 10^6$ for target #2, but during the simulation run it increases due to ionization (Fig. 3).

The purpose of choosing two targets is the following. For the thicker target the number of particles is large (several million) and the computations are spent primarily on pushing them. We call this "particle dominated" regime. The thinner target illustrates the opposite, "field dominated" regime, for which most computations are spend on computing the fields. Our goal is to investigate the system speed-up in the two regimes.

Simulations were done with a different number of processors. For each run the computation time was tracked and compared to a single-processor run, from which the computation speedup was evaluated. The computation speedup as a function of number of processors is plotted in Fig. 3c. The red dashed line is the theoretical maximum speed-up. In the "particle dominated" regime, the speed-up is very close to the maximum one. In the "field dominated" regime the speed-up is fairly good, about ~75−80 % of the maximum speed-up, somewhat lower compared to the thicker target. The reason is the increased rate of communications with the number of processors, which is more prominent if the number of particles is small. In both cases the speed-up is satisfactory, but we expect that for realistic cases it will be closer to the "particle dominated" regime since the parallelization targets computationally intensive problems, which arise primarily for "thick" targets ($L > 10\,\mu m$) with excessively large number of particles.

## 5. Conclusion

The implicit particle-in-cell algorithm developed previously has been parallelized using MPI (Message Passing Interface). The MPI implementation leads to a significant speed-up of the numerical code, especially in the case when computations are dominated by pushing particles. Future work will focus on inclusion of more detailed atomic physics and a binary collision model for small angle Coulomb scattering, as well as extension to three dimensions.

**Acknowledgement:**

This work was supported by the Naval Research Laboratory under the Base 6.1 program.

**Figure captions.**

Fig. 1. Computational domain, laser radiation, target, and domain decomposition (left). Local computational domain of processor "k" (right). The electric field values are defined on grid nodes, the magnetic field values are defined in cell center. The components of tensor $\hat{S}$ and vector $\delta \vec{j}$ are also defined on grid nodes.

Fig. 2. MPI parallelization sequence. The numbers correspond to the steps for computing the electromagnetic fields described in Section 3: (2) send one row of magnetic field values $\vec{H}_{0,j}$, $0 \leq j \leq m-1$ from processor $k+1$ to processor $k$; (3) apply boundary conditions at $x=0$ and $x=L_x$ to update $\vec{E}_{i,0}$ and $\vec{E}_{i,m}$, $0 \leq i \leq n$ on processor $k$; (4) add a source field at $x=0$ to update $\vec{E}_{i,0}$, $0 \leq i \leq n$ on processor $k$; (5) compute interior points for the electric field on processor $k$, $\vec{E}_{i,j}$, $1 \leq i \leq n, 1 \leq j \leq m$; (6) compute the electric field values $\vec{E}_{n,j}$, $0 \leq j \leq m$ belonging to the last row $n$ processor $k$; (7) send the last row of values $\vec{E}_{n,j}$, $0 \leq j \leq m$ from processor $k-1$ to processor $k$; (8) compute the magnetic field values $\vec{H}_{i,j}$, $1 \leq i \leq n-1, 0 \leq j \leq m-1$, for rows $i \geq 1$; (9) compute the magnetic field $\vec{H}_{0,j}$, $0 \leq j \leq m-1$ for row $i=0$.

Fig. 3 Time evolution of the number of computational particles in the "particle dominated" regime (a), and "field dominated" regime (b). Computation speedup versus number of processors for the two regimes (c). Squares: particle dominated regime, cycles: field dominated regime, red dotted line: maximum speedup.

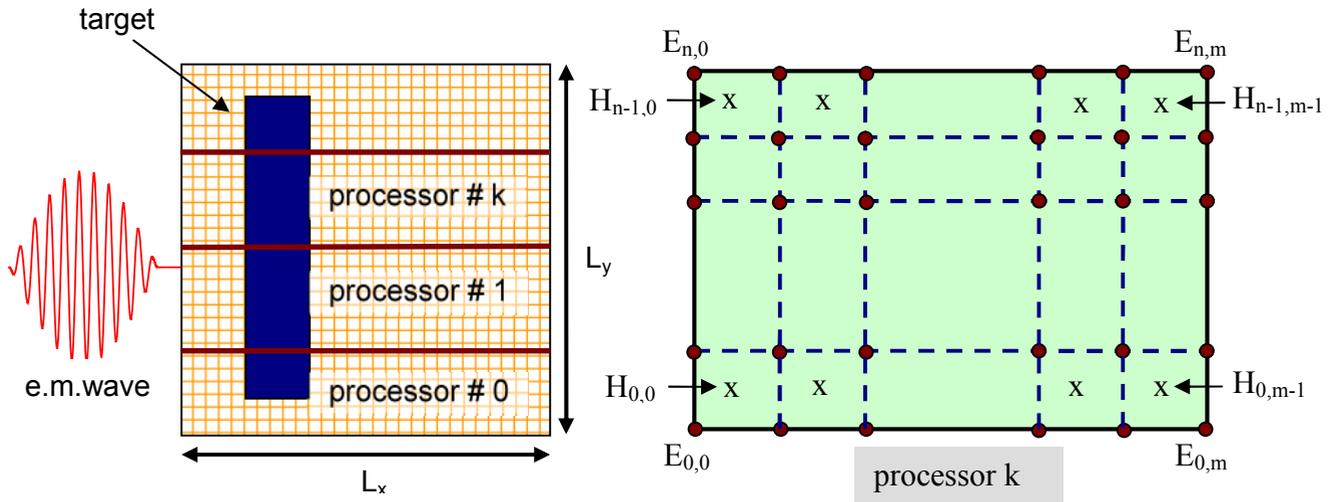

Fig. 1

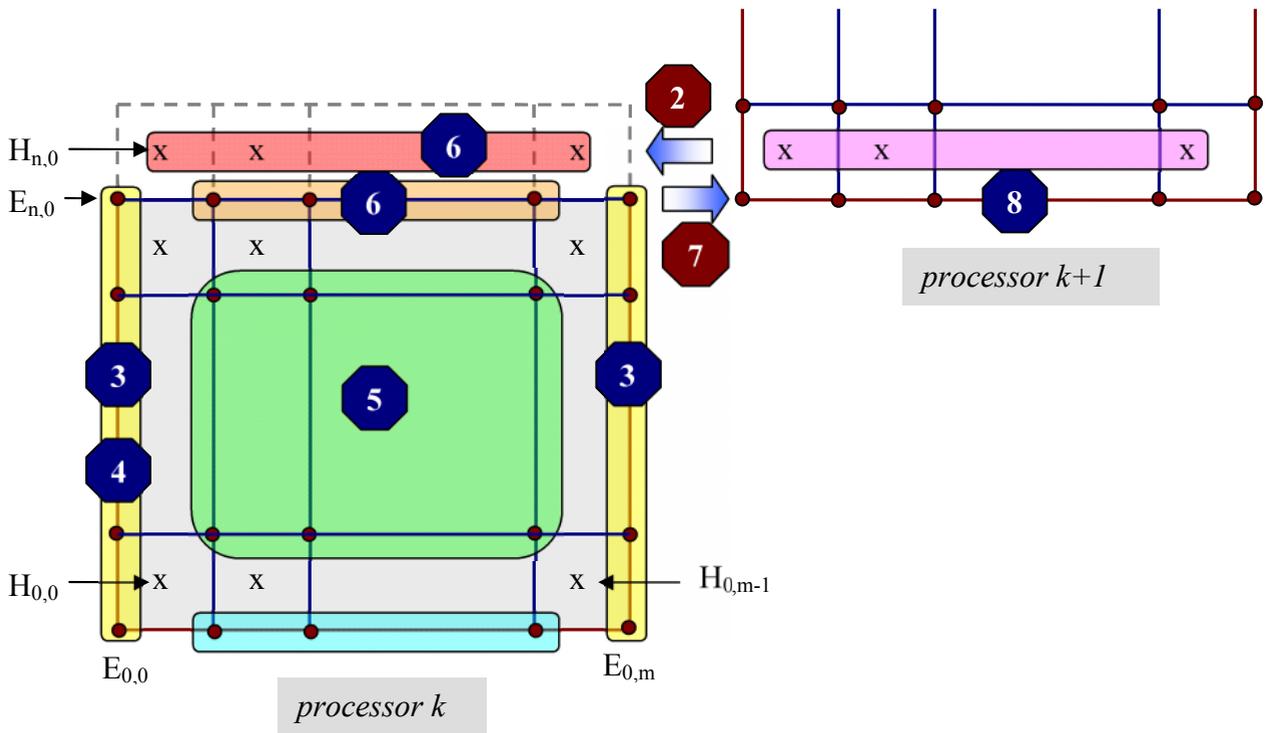

Fig. 2

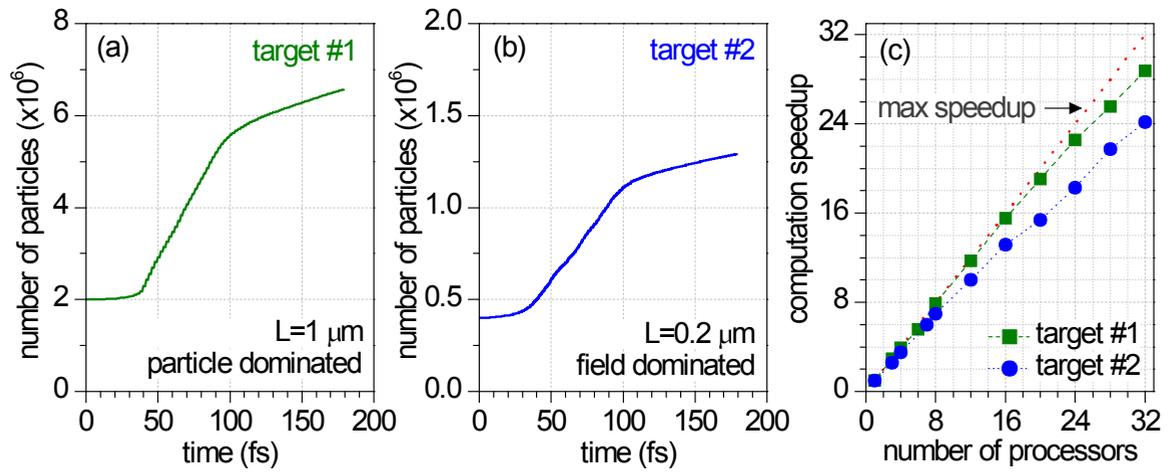

Fig. 3